\documentstyle[twoside,fleqn,epsf,espcrc2]{article}

\newcommand{\AmS}{{\protect\the\textfont2
  A\kern-.1667em\lower.5ex\hbox{M}\kern-.125emS}}

\title{Semileptonic Decays of Heavy Mesons:
       A Status Report\thanks{Work supported by DFG grant
Schi 257/1-4, Schi 257/3-2, EC contract CHRX-CT92-0051 and Mu 810/3.} }

\author{\underline{S. G\"usken}$^a$, K. Schilling$^{a,b}$
, G. Siegert$^a$ \vskip 0.3cm
        {$^a$ Physics Department, 
        University of Wuppertal, D-42097 Wuppertal, Germany \\
         $^b$ HRLZ, c/o KFA, D-52425 J\"ulich, Germany}}
       
\begin{document}

\begin{abstract}
We present intermediate results on our ongoing investigation concerning semileptonic 
decays of heavy pseudoscalar mesons into pseudoscalar and vector mesons.
The corresponding formfactors are evaluated
at several momenta and appropriate combinations of  four light and
four heavy quarks, which are chosen to allow for an extrapolation into the B Meson
region. In order to obtain clear groundstate signals we apply
gauge invariant ``Wuppertal'' smearing to the quarks.
The analysis is 
based on 32 quenched gauge configurations of size  $24^3 \times 64$ at $\beta=6.3$,
with Wilson fermions.
\end{abstract}

\maketitle

\section{INTRODUCTION}
The accurate determination of Kobayashi Maskawa (KM) matrix elements 
involving beauty quarks is an important
but also very challenging task to both experimental and 
theoretical physics.
On  the theoretical side nonpertubative methods are indispensable to calculate
the low energy QCD parts of the transition amplitudes. Although lattice
QCD is the method of choice in this context, it suffers from the fact
that the B meson region is beyond currently reachable lattice resolutions.
Direct lattice calculations at the B meson mass would be contaminated
by large discretization errors.

In the current project we are aiming at a high statistics
determination, with about 100 independent configurations, of the 
relevant weak formfactors, at a lattice 
resolution of $a^{-1} \simeq 3.2$ GeV. This resolution allows to push forward
into a mass region of $ (1 - 1.5) \times m_D$, with tolerable 
discretization errors.  We are running at four values of
the heavy quark mass $m_h$, with hopping parameters
$\kappa_h = 0.1200,01300,01350,0.1400$ and four values of the
light quark mass $m_l$, $\kappa_l = 0.1450,0.1490,0.1507,0.1511$. This
 covers the physical regions $ 0.8 m_{c} \leq m_h \leq 1.6 m_{c}$ 
and $ 0.8 m_s \leq m_l \leq 3 m_s$ respectively, providing  us with
sufficient  lever arms for the extrapolations to $m_u$ and $m_b$.

In this status report, we present the results of an intermediate
analysis, based on 32 configurations.

\section{LATTICE SIGNALS}
We consider the formfactors parametrizing the decay of pseudoscalar mesons,
$PS \rightarrow PS'$,
 $f^+(q^2),\; f^0(q^2)$, and the corresponding decays into vector
mesons, $PS \rightarrow V$, $V(q^2),\; A_1(q^2),\; A_2(q^2),\; A(q^2)$. 
They  can be extracted from the groundstate properties
of appropriate ratios $R(t)$ of
two- and three-point correlation functions\cite{Lubicz,Abada2}.
The three-point correlators
are calculated using the well known insertion technique\cite{Crisa}.

In our setup the initial meson is
located at timeslice
$t_i = 32$ and carries no momentum, whereas the final meson is placed at
timeslice $t_f = 1$ and is
 furnished with altogether 11 lattice momenta,
i.e. all combinations possible in the range 
$0 \leq |\vec{p}_f| \leq 2 p_{min}$, $p_{min}=2\pi/24a$.
The location in time  of the effective elektroweak interaction 
is varied between $t_i$ and $t_f$ and must be chosen finally in the
range of groundstate dominance, as signalled  by the existence of a plateau.

In order to enhance groundstate signals we have applied Wuppertal
``Gaussian'' smearing\cite{smearing} to all boundstate quarks,  with an average radius
of $\sqrt{\langle r^2 \rangle} \simeq 5a$.
We obtain clear plateaus  for all relevant ratios $R(t)$ with momenta
$0 \leq |p_f| \leq \sqrt{2}\times p_{min}$
in the range $10 \leq t \leq 20$. For higher momenta the signals start
getting noisy and a safe identification of the plateau region becomes
increasingly difficult.
\section{$q^2$ Dependence}
In order to extract the KM matrix elements from experimental measurements
of decay widths and branching ratios one conveniently uses the values of the formfactors
at $q^2 = 0$. A natural guide for the extrapolation in $q^2$ is the pole
dominance hypothesis
\begin{equation}
F(q^2) = \frac{F(0)}{1-\frac{q^2}{m^2_t}} \; .
\label{q2_dep}
\end{equation}
$F$ denotes generically the  formfactors given above and  $m_t$ is the mass 
of the lightest meson exchanged in the $t$ channel.
The ansatz can be checked by the ``high momentum transfer'' data, at 
$|\vec{p}| = \sqrt{3}p_{min}$ and $2 p_{min}$.
\begin{figure}[htb]
\vskip -0.7cm
\epsfxsize=7.0cm
\epsfbox{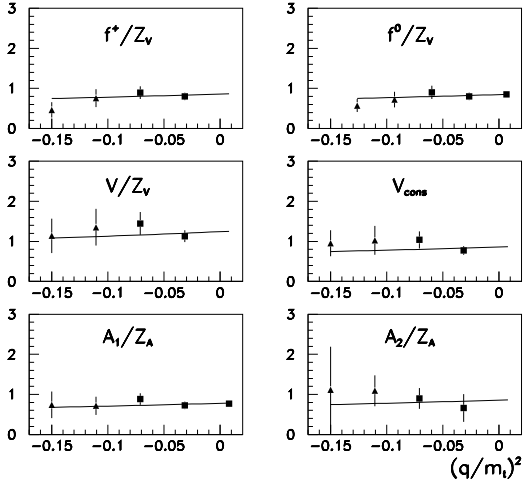}
\vskip -0.8 cm
\caption[a]{\label{pole_dominance}
\it{$q^2$ dependence of formfactors.The mass of the heavy initial
and the heavy final quark corresponds to $\kappa_i = 0.120$ and
$\kappa_f=0.135$ respectively. The mass of the light spectator quark is
$\kappa_{spec}=0.1490$ and the mass $m_t$ was extracted from appropriate
2-point correlators. The triangles denote the ``high momentum'' results.}} 
\end{figure}
In fig.\ref{pole_dominance} we show
\vskip -0.5cm
 the $q^2$ dependence of the formfactors
for a given  combination of quark masses. The solid line corresponds to
a fit to eq.\ref{q2_dep}, where we have included only the most reliable
low momentum results. 
We find reasonable consistency of all data
with the pole dominance hypothesis. We would like to  emphasize  that
the extrapolation to  $q^2=0$  depends only weakly  on the details of the
ansatz as it is  short ranged.    
\section{$D$ Meson Decays}
The charm quark region is directly accessed by our calculation. After
extrapolation in $q^2$ and in the light quark mass  $m_l \rightarrow m_u$ and 
$m_l \rightarrow m_s$ we find 
\begin{figure}[htb]
\vskip -1.1cm
\epsfxsize=7.0cm
\epsfbox{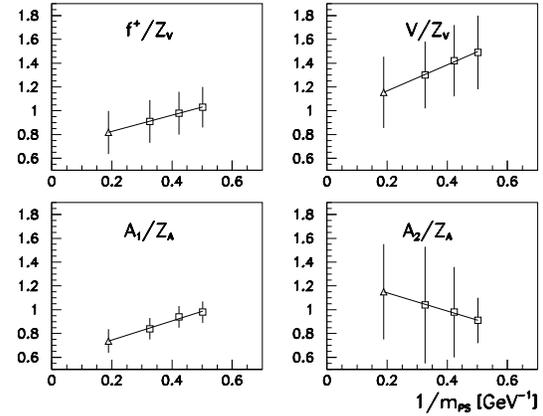}
\vskip -1.5 cm
\caption[a]{\label{isgur_plot}\it{$1/M_{ps}$ dependence of formfactors
 in the heavy quark regime.
The light quark mass is extrapolated to the chiral point and the mass
of the final heavy quark corresponds to $m_c$. The triangle represents
the result of the fit at $M_B$.}} 
\end{figure}
\vskip -1.25cm
\begin{eqnarray}
\lefteqn{D \rightarrow Kl\nu,K^*l\nu:} \nonumber \\
&&\!\!\!  f^+(0)/Z_V = 1.02(20) \quad f^0(0)/Z_V = 1.01(8)  \nonumber \\
 &&\!\!\! V(0)/Z_V = 1.41(45) \nonumber  \\
 &&\!\!\! A_1(0)/Z_A = 0.74(10) \quad A_2(0)/Z_A = 0.70(47) \nonumber 
\end{eqnarray}
and
\begin{eqnarray}
\lefteqn{D \rightarrow \pi l\nu,\rho l\nu:} \nonumber \\
&& \!\!\! f^+(0)/Z_V = 1.07(22) \quad f^0(0)/Z_V = 1.04(8) \nonumber \\
&& \!\!\! V(0)/Z_V = 1.46(48) \nonumber \\
&& \!\!\! A_1(0)/Z_A = 0.78(11) \quad A_2(0)/Z_A = 0.73(52) \; .\nonumber 
\end{eqnarray}
  A discussion of the issue
of the  renormalization constants of the vector-
and axial vector-currents, $Z_V$ and $Z_A$,
 will follow in section 6.
\section{$B$ Meson Decays}
The $B$ meson can be accessed only by extrapolation in the heavy quark 
mass. The heavy quark effective theory predicts\cite{isgur}
\begin{equation}
F(M_{ps}) = c + \frac{d}{M_{ps}} \; ,
\label{isgur_wise} 
\end{equation}
if both the initial and the final quark are ``heavy''. In fig.\ref{isgur_plot}
we display our heavy quark results together with a fit
 to eq.\ref{isgur_wise}. Obviously the data, within the present
 statistics,
 is well described by 
eq.\ref{isgur_wise}, and we obtain for
the transitions
\begin{eqnarray}
\lefteqn{B \rightarrow Dl\nu,D^*l\nu:} \nonumber \\
&& \!\!\! f^+(0)/Z_V = 0.82(20) \quad V(0)/Z_V = 1.15(30) \nonumber \\
&& \!\!\! A_1(0)/Z_A = 0.74(20) \quad A_2(0)/Z_A = 1.15(60) \; .\nonumber 
\end{eqnarray}
\begin{figure}[htb]
\vskip -1.2cm
\epsfxsize=7.0cm
\epsfbox{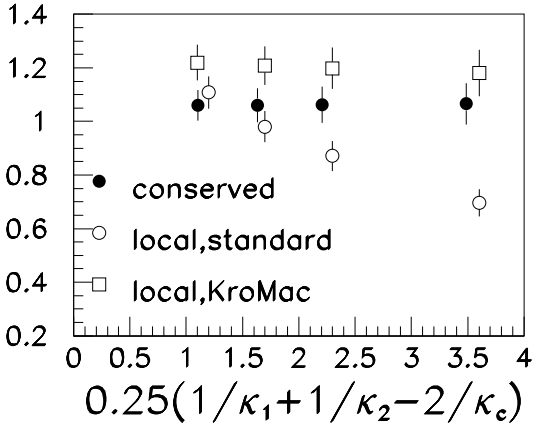}
\vskip -0.3 cm
\caption[a]{\label{PSVKPS}
\it{Mass dependence of $M_{loc}$ and $M_{cons}$ for degenerate
pseudoscalar states. 
The mass of the light spectator quark is $\kappa_{spec}=0.1490$.}}
\end{figure}
\section{Mass Dependence of $Z_V$}
In order to convert the above lattice data into continuum results, we need the 
renormalization constants $Z_V$ and $Z_A$ of the vector and axialvector currents.
They can be calculated in perturbation theory, but  the estimation
of the systematic error is difficult in this case.

 Therefore it is 
important to gain information about the renormalization constants directly
from the nonperturbative data.

One year ago C.W. Bernard \cite{Bernard} argued that the renormalization constant
of the local vector current $Z_V$ depends strongly on the mass of the
quarks involved. Following his arguments this undesired mass
 dependence can be removed if 
the (standard) $\sqrt{2\kappa}$ normalization of the quark fields is
replaced by the Kronfeld-Mackenzie(KroMac) \cite{KroMac} prescription.

As a first step to check this issue\footnote{see also \cite{Vladikas}.} we 
display in fig.\ref{PSVKPS} the mass dependence of the matrix element
$M_{loc}=(\langle PS|V^{loc}_0|PS\rangle/Z_V)_{q^2=0}$
 with the standard and with the KroMac normalization.
We have also included 
the result for the conserved current 
$M_{cons}=\langle PS|V^{cons}_0|PS\rangle_{q^2=0}$, which is
expected to be 1 in the standard normalization because of the corresponding
Ward identity.
 
Obviously the data favor the KroMac normalization for the local current,  
as for this choice $Z_V = M_{cons}/M_{loc}$ becomes mass independent. 

In a forthcoming paper we will extend the above considerations
to the space components
of the vector and axialvector currents.
\vskip 0.2cm
{\bf Acknowledgements}: We thank R. Gupta, A.S. Kronfeld and A. Vladikas
for useful discussions related to section 6.


\begin{thebibliography}{99}
\bibitem{Lubicz}{V. Lubicz et al.; Nucl. Phys. B356 (1991)301.}
\bibitem{Abada2}{A. Abada et al.; Nucl. Phys. B416 (1994)675.}
\bibitem{Crisa}{M. Crisafulli et al.; Phys. Lett. B223 (1989)90.}
\bibitem{smearing}{C. Alexandrou et al; Nucl. Phys. B414(1994)815.}
\bibitem{isgur}{N. Isgur and M.B. Wise; Phys. Rev. D42 (1990)2388.}
\bibitem{Bernard}{C.W. Bernard; Nucl. Phys. B(Proc. Suppl.) 34(1994)47.}
\bibitem{KroMac}{A.S. Kronfeld, Nucl. Phys. B(Proc. Suppl.) 30(1993)445; P.B.
Mackenzie, Nucl. Phys. B(Proc. Suppl.) 30(1993)35.}
\bibitem{Vladikas}{A. Vladikas et al.; these proceedings.}
\end{thebibliography}
\end{document}